\documentclass[aps,pre,unsortedaddress,reprint]{revtex4-1}
\usepackage{graphicx}
\usepackage{amsmath}
\usepackage{amssymb}

\begin{document}
\title{Adsorption isotherms for charged nanoparticles}

\author{Alexandre P. dos Santos}
\email{alexandre.pereira@ufrgs.br}
\affiliation{Instituto de F\'isica, Universidade Federal do Rio Grande do Sul, Caixa Postal 15051, CEP 91501-970, Porto Alegre, RS, Brazil}

\author{Amin Bakhshandeh}
\email{bakhshandeh.amin@gmail.com}
\affiliation{Instituto de F\'isica, Universidade Federal do Rio Grande do Sul, Caixa Postal 15051, CEP 91501-970, Porto Alegre, RS, Brazil}

\author{Alexandre Diehl}
\email{diehl@ufpel.edu.br}
\affiliation{Departamento de F\'\i sica, Instituto de F\'\i sica e Matem\'atica, Universidade Federal de Pelotas, Caixa Postal 354, CEP 96010-900, Pelotas, RS, Brazil}

\author{Yan Levin}
\email{levin@if.ufrgs.br}
\affiliation{Instituto de F\'isica, Universidade Federal do Rio Grande do Sul, Caixa Postal 15051, CEP 91501-970, Porto Alegre, RS, Brazil}

\begin{abstract}
We present theory and simulations which allow us to quantitatively calculate the amount  of surface adsorption excess of charged nanoparticles to a charged surface.  The theory is very accurate for weakly charged nanoparticles and can be used at physiological concentrations of salt.  We have also developed an efficient simulation algorithm which can be used for dilute suspensions of nanoparticles of any charge, even at very large salt concentrations. With the help of the new simulation method, we are able to efficiently calculate  the adsorption isotherms of highly charged nanoparticles in suspensions containing multivalent ions, for which there are no accurate theoretical methods available. 

\end{abstract}
 
\maketitle

\section{Introduction}

The interaction between lipid membranes, DNA, electrodes, and other charged surfaces with nanoparticles is of fundamental importance in biochemistry, biophysics, and diagnostic medicine. 
It is well known that salt can modify significantly the interaction between biomolecules in aqueous suspensions, affecting their stability~\cite{BrTo96,VeKa98,GoCa98,HaLo00,Li01,Sedu03,LiLu04b,BoGu15,MePr15,PlHa15,DiLi15}. The Derjaguin-Landau-Verwey-Overbeek~(DLVO) theory~\cite{Ru89} attributes the stability of suspensions to the competition between electrostatic and dispersive, van der Waals~(vdW), forces. Electrostatic repulsion between colloidal particles prevents them from coming into a close contact at which strong dispersion forces 
can make the particles  stick together, resulting in flocculation and precipitation. The  
vdW attraction is very short-ranged and is only weakly affected by the presence of electrolyte.
On the other hand, the Coulomb repulsion between like charged particles is strongly susceptible 
to the presence of electrolyte, which screens the electrostatic interactions.
The DLVO theory provides a qualitative understanding of stability of colloidal systems in suspensions containing 1:1 electrolyte.  The theory, however, is not able to account for either the ionic specificity (Hofmeister effect)~\cite{Ba96,LoJo03,ScHo10,DoLe11,Zh16}, like-charge attraction~\cite{Pa80,Gu84,MaNa08,TrSa11}, or the reversal of the electrophoretic mobility often observed in suspensions containing multivalent ions~\cite{LoGo99,KuKr12,SeRa13}.
In this paper, we will explore the interaction between  nanoparticles and charged surfaces.
Our goal is to quantitatively calculate the adsorption isotherms for  dilute 
suspensions of nanoparticles in solutions containing large -- physiological concentrations -- of electrolyte. 

When studying Coulomb systems the starting point is often the Poisson-Boltzmann~(PB) equation.
Indeed, it has been observed that PB equation can very accurately describe the density
profiles of monovalent ions near a charged wall.  However, since the PB equation does not
take into account either electrostatic correlations or steric repulsion between ions it
is bound to fail if used to calculate the adsorption isotherms of 
charged nanoparticles near a charged wall~\cite{Le02}.  Nevertheless, we will show that a very simple
modification of PB equation can extend its validity to study an important class of weakly charged nanoparticles,  allowing us to quantitatively calculate their adsorption isotherms.
For a more strongly charged nanoparticles, or if solution contains multivalent ions, we will present
a simulation method which allows us to obtain adsorption isotherms at infinite dilution of 
nanoparticles, which are often of great practical interest.

The paper is organized as follows: in Section~\ref{mt}, we introduce a modified PB (mPB) equation which allows us to accurately calculate the density profiles of weakly 
charged nanoparticles near a charged surface. In Section~\ref{simul} we present an  efficient Monte Carlo~(MC) simulation method which can be used to obtain the adsorption isotherms for very dilute suspensions of nanoparticles at large salt concentrations.  In section~\ref{res}, we compare the theory with the simulations and discuss suspensions containing multivalent ions. In Section~\ref{conc}, conclusions of the present work are presented.

\section{Model and Theory}\label{mt}

Consider a spherical colloidal particle of radius $a$ and charge $Z$ in an electrolyte solution.  If  $Z$ is not too large, there is no counterion condensation, and 
the electrostatic potential produced by a colloidal particle inside a 1:1 electrolyte
solution can be found analytically by solving the linearized PB equation~\cite{DeHu23,Le02}, 
\begin{equation}\label{eq1}
\beta\phi(r)=\lambda_B Z \left( \dfrac{e^{\kappa_e a}}{1+\kappa_e a}\right) \dfrac{e^{-\kappa_e r}}{r} \ ,
\end{equation}
where $r$ is the distance from the center of the nanoparticle, $\kappa_e=\sqrt{8\pi \lambda_B \rho_S}$ is the bulk inverse Debye length, $\rho_S$ is the salt concentration, $\lambda_B=q^2/\epsilon k_BT$ is the Bjerrum length, $q$ is the proton charge and $\epsilon$ is the dielectric constant of the medium. We observe that the electrostatic potential in Eq. (\ref{eq1}) 
is identical to the potential produced by a point charge of
\begin{equation}\label{zeff}
Z_{eff}=Z \left(\dfrac{e^{\kappa_e a}}{1+\kappa_e a}\right) \,.
\end{equation}
Note that $Z_{eff}$ can be significantly larger than the bare charge $Z$.  The interpretation of
this curious result is that if we want to replace a finite sized colloidal particle by a point particle and require that this point particle produces the same electric field at a distance $r\gg a$, the charge of the point particle must be larger than the bare colloidal charge in order to account for the absence of screening inside the colloidal core~\cite{LiLe94,FiLe94}.  This suggests that in the absence
of the counterion condensation, the system of weakly charged nanoparticles can be mapped onto a system of point particles with an effective charge given by Eq. (\ref{zeff}).

The dispersion interactions between nanoparticles and the surface can be taken into account using the Hamaker potential which can be written as~\cite{Ru89},
\begin{eqnarray}\label{van}
U_v(z)=-\dfrac{A}{6}[(\dfrac{{\bar a}}{(z-{\bar a})} + \dfrac{2 {\bar a}}{[4{\bar a} + 2(z-{\bar a})]} + \\
\log[\dfrac{2(z-{\bar a})}{4{\bar a} + 2(z-{\bar a})}])] \ , \nonumber
\end{eqnarray}
where $A$ is the Hamaker constant, set to $1.3\times10^{-20}$J, $\approx3.15~k_BT$ corresponding to polystyrene-polystyrene interaction in water at room temperature~\cite{Israel},  and ${\bar a}=a-4$\AA\ is the vdW radius of the nanoparticle (radius minus the hydration layer)~\cite{Bost}. We expect that the pairwise additive approximation on which Hamaker potential is based will break down at short separations, where we would need to use the Lifshitz theory~\cite{Pa06}.  In the present paper we will neglect this non-additive short distance effects.

Now, suppose that we have a dilute suspension of charged nanoparticles inside a 1:1 electrolyte solution.
If an oppositely charged surface is introduced into solution some of the particles will become adsorbed to it.
The surface adsorption excess can be defined as
\begin{equation}\label{ads}
\Gamma = \int_{0}^{\infty}[\rho(z)-\rho(\infty)]dz \,,
\end{equation}
where $\rho(z)$ is the number density of nanoparticles at a distance $z$ from the surface, and 
$\rho(\infty)=\rho_B$ is the bulk nanoparticle concentration.

To calculate the surface adsorption excess we need to know the 
density profile of nanoparticles $\rho(z)$.  It is well known that for weekly charged small ions,
Poisson-Boltzmann theory is very accurate, however, it fails 
for large or strongly charged ions~\cite{Le02}.  On the other hand, from the argument above we saw that for nanoparticles which are not too strongly charged, the effect of hardcore can be taken into account by simply renormalizing the colloidal charge.  In this sense, we can map weakly charged nanoparticles onto point particles with an effective charge.  Since the PB equation works very well for point-like ions, we expect that it will also work reasonably well for our weakly charged  nanoparticles which are mapped onto point-like particles with an effective charge~\cite{KaNa10,DiLi15}.  Note that in this formalism, the electrostatic correlations between the nanoparticles and the ions are taken into account through the charge renormalization.  A modified PB (mPB) equation for this system can then be written as 
\begin{eqnarray}\label{pb}
\nabla^2 \phi(z)&=&-\dfrac{4\pi q }{\epsilon_w} \left[\sigma + q\rho_+(z) - q\rho_+(z)+Z\rho(z)\right]  \ ,  \\
\rho(z)&=& \rho_B e^{- \beta Z_{eff}(z) \phi(z)-\beta U_v(z)-\beta U_e(z)} \ , \nonumber  \\
\rho_+(z)&=& \rho_s e^{-\beta q \phi(z)} \ , \nonumber \\
\rho_-(z)&=& \rho_s  e^{\beta q \phi(z)} \ , \nonumber
\end{eqnarray}
where $z$ is the distance from the charged wall, $\phi(z)$ is the mean electrostatic potential, $\rho(z)$, $\rho_+(z)$, $\rho_-(z)$, are the  density profiles of  nanoparticles, cations, and anions, respectively, and $\rho_s$ is the bulk concentration of salt. The hardcore potential $U_e(z)$ prohibits the centers of nanoparticles from coming nearer than a distance $a$ to the surface. The vdW interaction between the nanoparticles and the surface is given by Eq.~\ref{van}. The effective charge which appears in the mPB, Eq.~\ref{pb}, is calculated using the local density approximation
\begin{equation}
Z_{eff}=Z \left(\dfrac{e^{\kappa(z) a}}{1+\kappa(z) a}\right) \ ,
\end{equation}
where
\begin{equation}\label{kappa}
\kappa(z)=\sqrt{4\pi \lambda_B [\rho_+(z)+\rho_-(z)]}
\end{equation}
is the local inverse Debye length. This is similar to the well known WKB approximation~\cite{Po89}. The Bjerrum length is set to $7.2~$\AA, value for water at room temperature. The mPB equation can be solved numerically using Picard iteration method.  To check the accuracy of the mPB equation we compare its predictions with the results of Monte Carlo simulations.

\section{Monte Carlo Simulations}\label{simul}

In order to accurately construct the nanoparticle density profile for dilute suspensions at physiological concentrations of salt requires a very large simulation cell containing many ions. The long range Coulomb force prevents us from using simple periodic boundary conditions, requiring more sophisticated Ewald summation methods which are computationally very expensive. Furthermore presence of many salt ions results in very low MC acceptance rates,  requiring introduction of cluster~\cite{LoLi99} or inversion moves~\cite{LiLu04a}, leading to additional complications. To overcome these difficulties we have developed a new approach for calculating adsorption isotherms of dilute suspensions using MC simulations. Our algorithm is based on the fundamental observation that the density profile of nanoparticles 
can be written as
\begin{equation}\label{prof}
\rho(z)=\rho_B e^{-\beta \omega(z)} \ ,
\end{equation}
where $\omega(z)$ is the potential of mean force. For very dilute suspension of nanoparticles in a solution containing large amount of salt, the interaction between nanoparticles can be ignored, so that the potential of mean force depends only on the surface charge density and the concentration of electrolyte.

The MC simulations are performed in a box of sides $L_x=L_y=218~$\AA\ and $L_z=5L_x$. The electrolyte is confined in $z$ direction between $z=0$ and $z=L=150~$\AA. A charged wall with a uniform surface charge density $\sigma=-0.03~\text{C}/\text{m}^2$ is located at $z=0$. A nanoparticle has charge $Z=5q$ and effective radius $a=20~$\AA, similar to lysozyme~\cite{Kuehner}, where $q$ is the proton charge, and is placed  at position $z$ and $x=0$ and $y=0$. We also consider $N_c=\left(\frac{89}{\alpha}-\frac{Z}{\alpha q}\right)/L_xL_y$ counterions of charge $~\alpha q$, where $\alpha$ is the ionic valence. Positive and negative ions from dissociation of $\alpha$:1 electrolyte are also present in the system. All ionic species have radius $2~$\AA. Water is treated as a uniform medium  of dielectric constant $\epsilon=80\epsilon_0$, where $\epsilon_0$ is the dielectric constant of vacuum.  The electrostatic interactions are calculated by summing over all the periodic replicas of the system using Ewald summation method, modified for slab geometry~\cite{YeBe99,PoDa14}. Here we adopt a recently introduced efficient simulation algorithm developed specifically for this geometry~\cite{DoGi16}.
The electrostatic energy of a periodically replicated system, containing $N$ charged particles, is
\begin{eqnarray}\label{ener}
U=\sum_{{\pmb k}\neq{\pmb 0}}^{\infty}\frac{2\pi}{\epsilon_w V |{\pmb k}|^2}\exp{[-\frac{|{\pmb k}|^2}{4\kappa_e^2}]}[A({\pmb k})^2+B({\pmb k})^2] + \nonumber \\
\frac{2\pi}{\epsilon_w V}[M_z^2-Q_tG_z]
+ \hspace{1cm} \nonumber \\
\dfrac{1}{2}\sum_{i \ne j}^N\text{q}_i\text{q}_j\frac{\text{erfc}(\kappa_e|{\pmb r}_i-{\pmb r}_j|)}{\epsilon_w |{\pmb r}_i-{\pmb r}_j|} - \nonumber \\
\dfrac{2\pi}{\epsilon_w}\sum_{i=1}^{N}\sigma z_i\text{q}_i \ , \hspace{1cm}
\end{eqnarray}
where
\begin{eqnarray}
A({\pmb k})=\sum_{i=1}^N \text{q}_i\text{cos}({\pmb k}\cdot{\pmb r}_i) \ , \nonumber \\
B({\pmb k})=-\sum_{i=1}^N \text{q}_i\text{sin}({\pmb k}\cdot{\pmb r}_i) \ , \nonumber \\
M_z=\sum_{i=1}^N \text{q}_i z_i \ , \nonumber \\
Q_t=\sum_{i=1}^N \text{q}_i \ , \nonumber \\
G_z=\sum_{i=1}^N \text{q}_i (z_i)^2 \ .
\end{eqnarray}
The k-vector is defined as $\pmb k=(2\pi\dfrac{n_x}{L_x},2\pi\dfrac{n_y}{L_y},2\pi\dfrac{n_z}{L_z})$, where $n's$ are integers. $V=L_xL_yL_z$ is the volume of the simulation box, $\kappa_e=4/L_x$ is the dumping parameter of the Ewald summation method, q$_i$ and ${\pmb r}_i$ are the charge and position of particle $i$, respectively. The MC simulations are performed using Metropolis algorithm with $10^7$ movements to achieve equilibrium and $100$ movements per particle to obtain uncorrelated states. The force profiles are obtained using $30000$ uncorrelated states.  To achieve convergence of the electrostatic energy we use around $400$ k-vectors. 

For a  nanoparticle fixed at a distance $z$ from the charged surface, we calculate the ensemble averaged electrostatic and entropic forces acting on the particle. The electrostatic force is given by
\begin{equation}
F_{elec} = - \left\langle\frac{\partial U}{\partial z} \right\rangle \ , 
\end{equation}
where $U$ is the electrostatic energy of the system~\cite{DoGi16}.

The entropic force is obtained using the approach of Wu~{\it et al.} which requires performing a virtual displacement of the nanoparticle and counting the number of overlaps with the ions of electrolyte~\cite{WuBr99}. It is given by
\begin{equation}
F_{ent}=\frac{<N^c>-<N^f>}{\beta \Delta R} ,
\end{equation}
where $N^c$ is the number of virtual overlaps between the colloid and the ions after a small displacement $\Delta R=0.5~$\AA~ that brings colloids and plate closer together (superscript $c$ stands for closer) and $N_f$ is the number of overlaps of the colloids and the ions after a displacement $\Delta R$ that moves the colloids and plate farther apart (superscript $f$ stands for farther).

After the force profile is calculated, the potential of mean force is obtained by integration
\begin{eqnarray}
\omega(z)=U_v(z)+\int_{z}^{\infty}\left[ F_{elec}(z')+F_{ent}(z') \right] dz' \ .
\end{eqnarray}
The great advantage of this method is that the calculation of force is easily parallelized by running it on different CPUs for each $z$.

\section{Results}\label{res}

\begin{figure}[h]
\begin{center}
\includegraphics[width=7.cm]{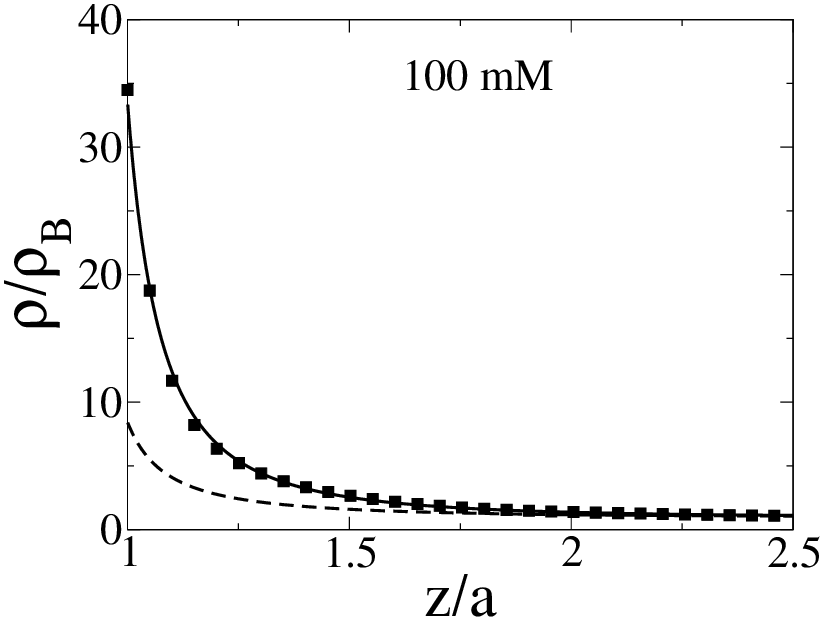}\vspace{0.2cm}
\includegraphics[width=7.cm]{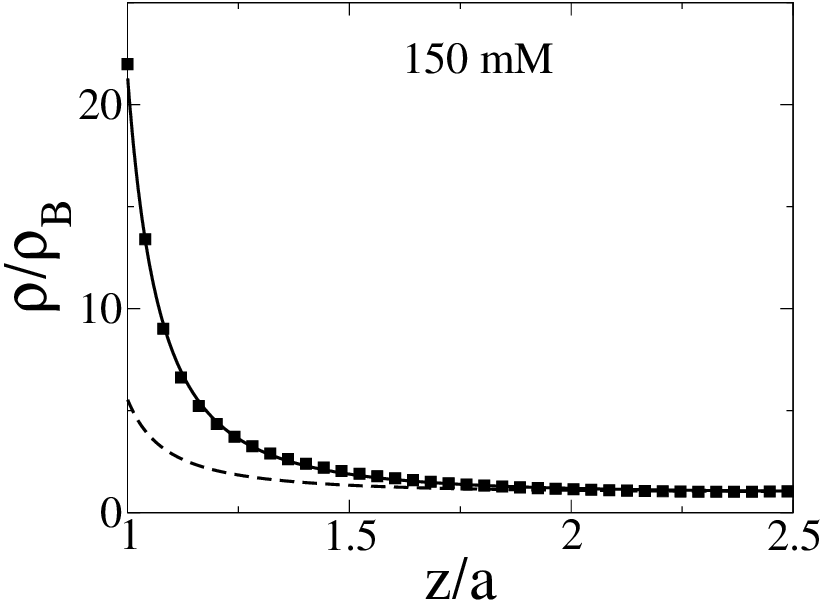}\vspace{0.2cm}
\includegraphics[width=7.cm]{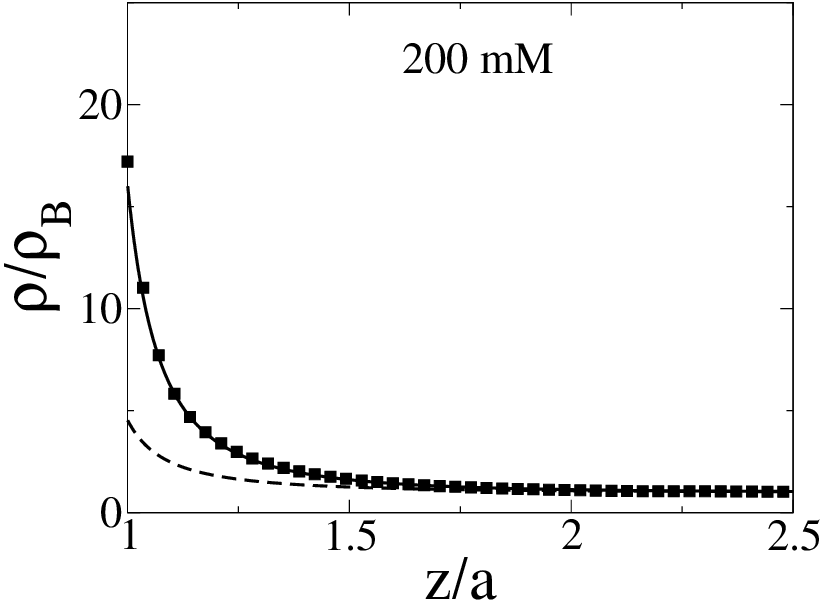}\vspace{0.2cm}
\end{center}
\caption{Density profiles of nanoparticles for salt concentrations $100$, $150$ and $200~$mM near a charged wall with $\sigma=-0.03~$C$/$m$^2$. Symbols are the results of MC simulations and solid lines are the predictions of the present theory. The dashed lines represent a solution of PB equation without taking into account charge renormalization. The bare nanoparticle charge is $Z=5q$ and radius is $a=20$\AA.}
\label{fig1}
\end{figure}

\begin{figure}[h]
\begin{center}
\includegraphics[width=7.cm]{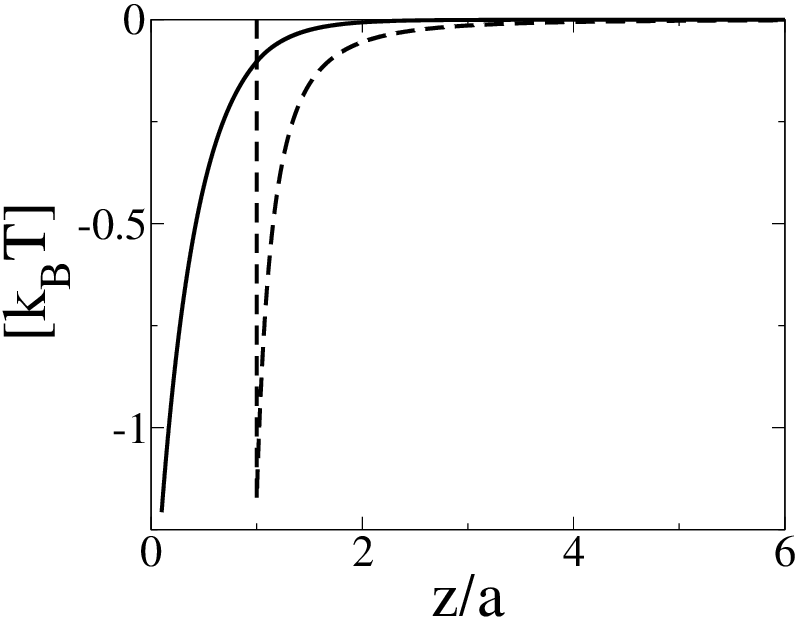}\vspace{0.5cm}
\end{center}
\caption{The scaled electrostatic potential, $\beta q \phi(z)$, solid line, and vdW plus hard core potential, $\beta U_v(z)+\beta U_e(z)$, dashed line. The parameters are the same as Fig.~\ref{fig1} for salt at $150~$mM.}
\label{fig2}
\end{figure}

In Fig.~\ref{fig1}, we plot the density profiles of nanoparticles
for different salt concentrations obtained using a numerical solution of 
Eq.~\ref{pb}. The agreement between simulations and theory is excellent.  In the same figure, the dashed lines
show the density profiles which are obtained if charge renormalization 
is not taken into account.  In this case we see a very strong deviation from the results of MC simulations.
The electrostatic and vdW plus hard core potentials are shown in Fig.~\ref{fig2} for a specific set of parameters.

\begin{figure}[t]
\begin{center}
\includegraphics[width=7.cm]{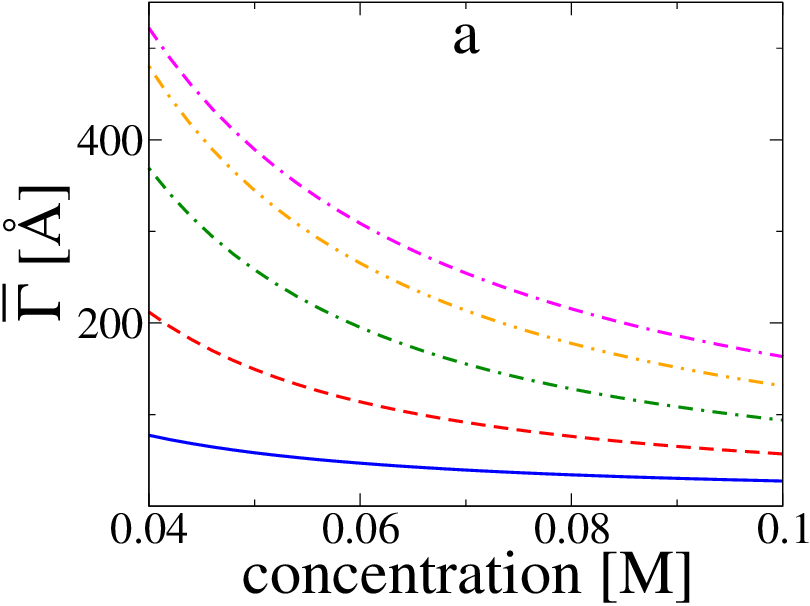}\vspace{0.25cm}
\includegraphics[width=7.cm]{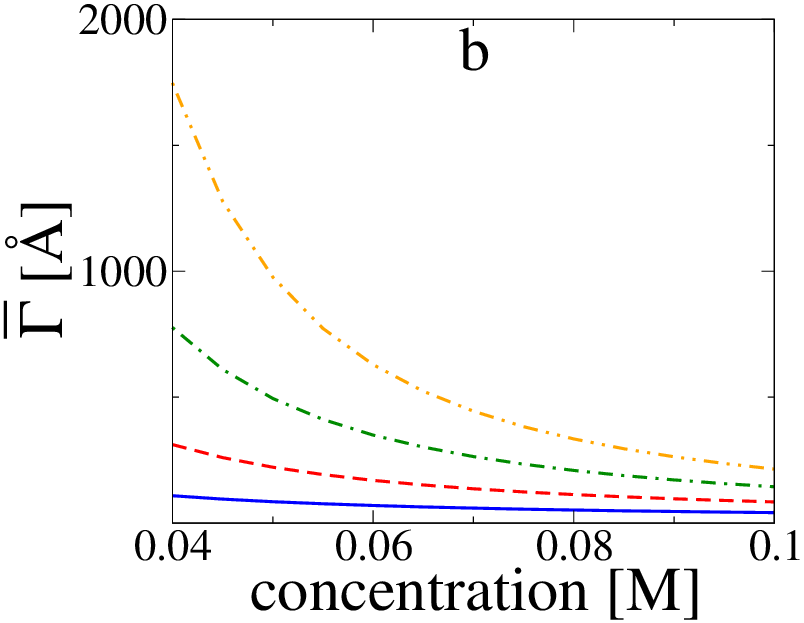}\vspace{0.25cm}
\end{center}
\caption{Surface excess vs. salt concentration calculated using mPB theory. In (a), $Z=5q$, while the surface charge densities are $\sigma=-0.01$, $-0.02$, $-0.03$, $-0.04$ and $-0.05~$C$/$m$^2$, from bottom to top, respectively. In (b), $\sigma=-0.03~$C$/$m$^2$ while the charge on the nanoparticle is $Z=3q$, $5q$, $7q$ and $9q$ from below to above, respectively. }
\label{fig3}
\end{figure}

The adsorption isotherms can be calculated by performing the integral in Eq.~\ref{ads}.
In Fig.~\ref{fig3} we plot the scaled adsorption isotherms, $\bar \Gamma = \Gamma/\rho_B$, as a function of salt concentration for various surface charge densities on the wall.  We see a very strong dependence of surface adsorption excess on the surface charge density at low salt concentrations.  For larger concentrations, Debye screening of electrostatic interactions leads to a much weaker dependence of surface adsorption excess on the wall surface charge density. 

For dilute colloidal suspensions, counterion condensation becomes important when $Z> 4 a (1+\kappa a)/ \lambda_B$ \cite{TrBo02,Le02}.  Indeed, when the nanoparticle charge exceeds this limit we see a significant  deviation between theory and simulations.  Furthermore, in this regime, we find that using more sophisticated theories to account for the counterion condensation and charge renormalization~\cite{AlCh84,TrLe04,CoLe09,DoDi09,CoDo12} is not sufficient to improve the agreement 
between theory and simulations.  Therefore, in order to calculate the adsorption isotherms of
strongly charged nanoparticles, or if suspension contains multivalent ions, we are forced to
rely on computer simulations which were discussed in  Section~\ref{simul}.

\begin{figure}[t]
\begin{center}
\includegraphics[width=7.cm]{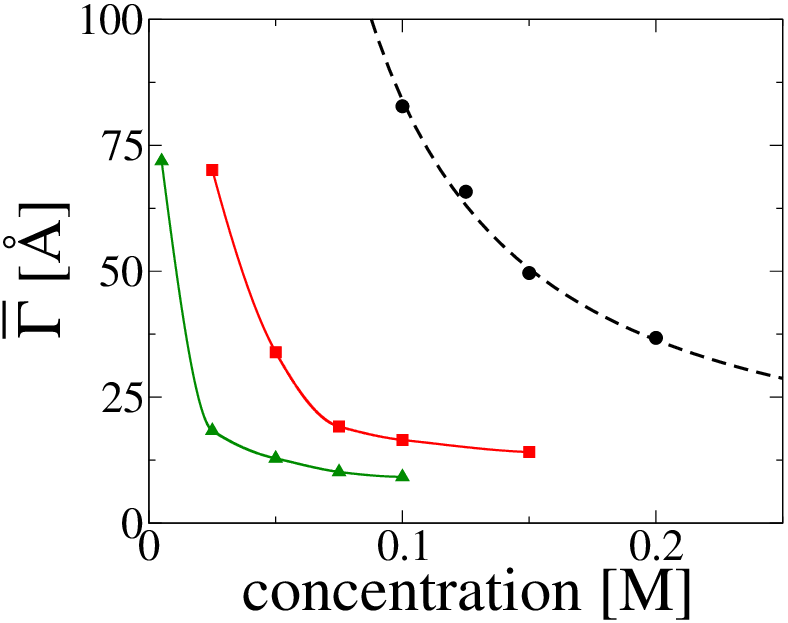}\vspace{0.5cm}
\end{center}
\caption{MC calculations for the rescaled surface excess vs. salt concentration for $Z=5q$. Circles represent monovalent salt, $\alpha=1$; squares, divalent counterions, $\alpha=2$; and triangles, trivalent counterions, $\alpha=3$. The solid lines are interpolations. The dashed line represents the theory presented in this paper.}
\label{fig4}
\end{figure}

In Fig.~\ref{fig4} we present  $\bar \Gamma$ for a dilute suspension, as a function of the added $\alpha$:1 electrolyte.
We see that screening of electrostatic interactions by electrolyte significantly reduces the nanoparticle adsorption. Furthermore, 
increasing cation valence,
$\alpha$, dramatically decreases the amount of adsorption,  see Fig.~\ref{fig4}. 
The figure also shows that for 1:1 electrolyte the adsorption isotherm 
calculated using mPB equation is in excellent agreement 
with the results of MC simulations. For more strongly charged nanoparticles, or in the presence of multivalent ions,  there are
no accurate theoretical methods available and one must rely on MC simulations. For example, in the case of highly charged nanoparticles, we observe more adsorption than is predicted by the  modified PB, Eq.~\ref{pb}. The mechanism of this attraction is both electrostatic and entropic in its origin. The counterions condensed onto nanoparticle  are repelled from the charged wall, shifting to the far side of the nanoparticle, leading to enhanced electrostatic and entropic attraction.

\section{Conclusions}\label{conc}

We have presented a theory which enables us to accurately calculate the density profiles and  adsorption isotherms of weakly charged nanoparticles. Both electrostatic and dispersion interactions between the nanoparticles and a charged surface are taken into account. The theory can be used even at large -- physiological -- concentrations of salt. However it fails for strong charged nanoparticles and strongly charged surfaces.  For such systems  we have developed an efficient MC algorithm which can be used to obtain both density profiles and the adsorption isotherms, which are of great practical importance in various applications. The simulations show that the counterion condensation near a strongly charged surface results in a short distance entropic repulsion  which is not properly captured by the mPB equation. The strength of this repulsion depends on the surface charge density and salt concentration.  For physiological salt 
concentrations used in the present paper the mPB equation remains accurate for surface 
charge densities up to $\sigma=-0.03~$C$/$m$^2$.  However for smaller salt concentration, the range of validity of the mPB equation increases.  For example, for $\approx60~$mM salt concentration, 
we find that mPB equation remains accurate for surfaces with $\sigma$ up to $-0.06~$C$/$m$^2$.
Finally, we note that the simulation approach developed in this paper can be easily applied to solutes of arbitrary shape and can also be extended to explicit solvent models.

\section{Acknowledgments}
This work was partially supported by the CAPES, CNPq, FAPERGS, INCT-FCx, and by the US-AFOSR under the grant FA9550-12-1-0438.

\bibliography{ref.bib}

\end{document}